\documentclass[fleqn,usenatbib]{stile/mnras}

\usepackage{txfonts}
\usepackage[T1]{fontenc}
\usepackage{ae,aecompl}

\usepackage{graphicx}

\usepackage{amsmath}
\usepackage{amsfonts}
\usepackage{amssymb}

\usepackage{xcolor}

\usepackage{hyperref}
\usepackage{footnote}

\usepackage{comment}

\usepackage{xfrac}
\def\nat{Nature}

\def\mnras{MNRAS}
\def\apj{ApJ}
\def\apjl{ApJL}
\def\apjs{ApJS}
\def\aap{A\&A}
\def\aaps{A\&A Supp.}

\def\aj{AJ}
\def\qjras{QJRAS}
\def\physrep{Phys. Rep.}
\def\apss{Ap\&SS}               % Astrophysics and Space Science
\def\pasa{Publ. Astr. Soc. Australia}

\newcommand{\quotes}[1]{``#1''}
\newcommand\code[1]{\textsc{\MakeLowercase{#1}}}

\def\mum{\mu{\rm m}}
\newcommand{\angstrom}{\textup{\AA}}
\def\pk{{\rm erg\,cm}^{-3}\,{\rm K}}
\def\skms{\sigma_{\rm kms}}
\def\tmyr{t_{\rm Myr}}
\def\msun{{\rm M}_{\odot}}
\def\lsun{{\rm L}_{\odot}}
\def\TdL{\langle T_d \rangle_L}
\def\TdM{\langle T_d \rangle_M}
\def\Tda{\langle T_d \rangle_a}

\def\gtsima{$\; \buildrel > \over \sim \;$}
\def\ltsima{$\; \buildrel < \over \sim \;$}
\def\gsim{\lower.5ex\hbox{\gtsima}}
\def\lsim{\lower.5ex\hbox{\ltsima}}

\def\be{\begin{equation}}
\def\ee{\end{equation}}

\def\HII{\hbox{H~$\scriptstyle\rm II $~}}

\def\CII{\hbox{[C~$\scriptstyle\rm II $]~}}

\usepackage{etoolbox}
\makeatletter
\newcount\c@additionalboxlevel
\newcount\c@maxboxlevel
\patchcmd\@combinedblfloats{\box\@outputbox}{%
  \stepcounter{additionalboxlevel}%
  \box\@outputbox
}{}{\errmessage{\noexpand\@combinedblfloats could not be patched}}

\AtBeginShipout{%
  \ifnum\value{additionalboxlevel}>\value{maxboxlevel}%
    \typeout{Warning: maxboxlevel might be too small, increase to %
      \the\value{additionalboxlevel}%
    }%
  \fi 
  \@whilenum\value{additionalboxlevel}<\value{maxboxlevel}\do{%
    \typeout{* Additional boxing of page `\thepage'}%
    \setbox\AtBeginShipoutBox=\hbox{\copy\AtBeginShipoutBox}%
    \stepcounter{additionalboxlevel}%
  }%
  \setcounter{additionalboxlevel}{0}%
}
\makeatother
\graphicspath{{plots/}}                 % folder(s) with plots

\begin{document}

% reference shortcuts
\defcitealias{F16}{F17}
\defcitealias{weingartner2001dust}{WD01}
\def\WD01{\citetalias{weingartner2001dust}}
\def\F17{\citetalias{F16}}
\def\C15{\citetalias{capak15}}

\title[Warm dust in high-z galaxies]{Warm dust in high-$z$ galaxies: origin and implications}

\author[Sommovigo et al.]{L. Sommovigo$^{1}$\thanks{\href{mailto:laura.sommovigo@sns.it}{laura.sommovigo@sns.it}},
A. Ferrara$^{1,2}$, 
A. Pallottini$^{1,3}$,
S. Carniani$^{1}$,
S. Gallerani$^{1}$,
\newauthor D. Decataldo$^{1}$\\
% list of institutions
$^{1}$ Scuola Normale Superiore, Piazza dei Cavalieri 7, I-56126 Pisa, Italy\\
$^{2}$ Kavli Institute for the Physics and Mathematics of the Universe (WPI), University of Tokyo, Kashiwa 277-8583, Japan\\
$^{3}$ Centro Fermi, Museo Storico della Fisica e Centro Studi e Ricerche \quotes{Enrico Fermi}, Piazza del Viminale 1, Roma, 00184, Italy
}

\label{firstpage}
\pagerange{\pageref{firstpage}--\pageref{lastpage}}

\maketitle

\begin{abstract}
ALMA observations have revealed the presence of dust in galaxies in the Epoch of Reionization (redshift $z>6$). However, the dust temperature, $T_d$, remains unconstrained, and this introduces large uncertainties, particularly in the dust mass determinations.
Using an analytical and physically-motivated model, we show that dust in high-$z$, star-forming giant molecular clouds (GMC), largely dominating the observed far-infrared luminosity, is warmer ($T_d \gsim 60\ \mathrm{K}$) than locally.
This is due to the more compact GMC structure induced by the higher gas pressure and turbulence characterising early galaxies.
The compactness also delays GMC dispersal by stellar feedback, thus $\sim 40\%$ of the total UV radiation emitted by newly born stars remains obscured.
A higher $T_d$ has additional implications: it (a) reduces the tension between local and high-$z$ IRX-$\beta$ relation, (b) alleviates the problem of the uncomfortably large dust masses deduced from observations of some EoR galaxies.
\end{abstract}


\section{Introduction}\label{intro}

Astrophysical dust is a crucial component of the interstellar medium (ISM) of galaxies both in the local and high redshift Universe \citep[e.g.][]{weingartner2001dust,2010...523A..85G,capak2011spectroscopy, riechers2013dust, weiss2013alma, laporte:2017apj}. Dust scatters and absorbs UV and optical light emitted by newly born stars, and re-emits it at far-infrared (FIR) and millimeter wavelengths as thermal radiation. Therefore, dust strongly impacts the observed flux and detectability of galaxies at these wavelengths \citep{calzetti2000dust,kriek2013dust}, despite it typically only accounts for less then few percent of the total ISM mass \citep{2007ApJ...657..810D}. 

Given the observed FIR spectra, Spectral Energy Distribution (SED) fitting techniques ideally allow us to infer the dust temperature, dust mass, total infrared luminosity, and obscured Star Formation Rate (SFR) of a galaxy \citep{sedfit_review,2014PhR...541...45C}. This is more common at $z<5$, thanks to both Herschel coverage at FIR wavelengths, and (sub)mm coverage from ground-based facilities, such as SCUBA, the Atacama Large Millimeter Array (ALMA), and the Northern Extended Millimeter Array (NOEMA) \citep[e.g.][]{sedfit_review,Kirkpatrick_2015}. 

For high-$z$ galaxies ($z>5$), deep FIR observations have only recently become possible thanks to the high sensitivity of millimeter interferometers such as ALMA and NOEMA \citep[e.g.][]{capak15,Bouwens16,willott2015star,laporte:2017apj,barisic2017dust,bowler2018obscured}. Nevertheless, to date, ALMA observations put only loose constraints on the SED shape of \quotes{normal} (i.e. {main-sequence}) galaxies in the Epoch of Reionization (EoR). This is because most of the current ALMA programs have observed the dust continuum emission only in a single (or two, in a few cases) ALMA band \citep[e.g.][]{Bouwens16,barisic2017dust,bowler2018obscured,2019MNRAS.487L..81L}. As a consequence, SED fitting techniques used at lower redshifts are not reliably applicable. Instead, given the broadband fluxes in the available ALMA band(s), a \quotes{dust temperature} must be assumed\footnote{We underline here that the \quotes{dust temperature} adopted for estimating $L_{\rm FIR}$ depends on the assumed dust SED functional shape, and does not necessarily reflect the physical dust temperature \citep[e.g. ][]{casey12,casey2018,2019MNRAS.tmp.2072L}} together with a functional shape of the dust SED in order to extrapolate the integrated FIR luminosity, $L_{\rm FIR}$ \citep[][]{capak15,Bouwens16,barisic2017dust,bowler2018obscured}. This SED sampling problem might be solved in the future thanks to proposed instruments such as SPICA \citep{spinoglio:2017,egami:2018}, featuring mid-infrared capabilities, and/or more extensive ALMA observations in other bands \cite{faisst20}. For the moment, as a result of the uncertainties in the assumed dust temperature, the many derived properties of high-$z$ galaxies are poorly constrained.

For instance, in the last years there has been tension regarding the surprisingly low Infrared excess (IRX) of high-$z$ galaxies, defined as ${\rm IRX} \equiv L_{\rm FIR} / L_{\rm UV}$, where $L_{\rm UV}$ is the rest-frame luminosity at $1600\ \mathrm{\angstrom}$ \citep[e.g.][]{1994calzetti,1997calzetti,meurer1999dust}. In particular, the relation between the IRX and the UV spectral slope $\beta$ at $z>5$ has been a matter of debate. This relation is particularly important since it is used to correct for dust obscuration in UV-selected galaxies at $z>5$. Interestingly, high-$z$ galaxies have been shown to have on average smaller IRX values with respect to local analogues, barred the relatively large scatter in the data  \citep{2014ApJ...796...95C,capak15,Bouwens16,2020arXiv200410760F}. Warmer dust would imply a higher FIR luminosity and IRX parameter, which would largely reconcile tension between the IRX-$\beta$ relationship of the local and the high-$z$ galaxies \citep{Bouwens16,faisst2017high,behrens2018dusty}.

Additionally, uncertainties in the dust temperature result in unreliable dust mass estimates. For instance, warmer dust can produce the same observed FIR flux without implying uncomfortably large dust mass, which are in tension with dust formation timescale constraints \citep[see e.g. ][]{lesniewska2019dust}. This issue clearly emerges, for example, in the analysis of the galaxy MACS0416Y1 at redshift $z=8.31$, for which \cite{bakx20} obtained an upper limit to the continuum flux at $160\ \mathrm{\mu m}$ (rest frame). Taken together with the previous continuum emission measurement at $90\ \mathrm{\mu m}$ \citealt{2019ApJ...874...27T}), \cite{bakx20} conclude that the dust temperature is $>80\ \mathrm{K}$. As a consequence, the dust mass estimate is reduced by a ten-fold with respect to the value deduced by \cite{2019ApJ...874...27T} ( which corresponds to $3-8\ \times 10^{6}\ \mathrm{M_{\odot}}$), obtained assuming $T_d = 40-50\ \mathrm{K}$ (see Sec. \ref{dustmass} for a detailed discussion).

In this context, theoretical studies are a crucial complementary tool in order to interpret high redshift observations. For instance, zoom-in simulations have  been  used  to  capture the internal properties and dynamics of galaxies in their cosmological environment, by consistently following the star formation history and feedback  processes driving the chemical evolution of high-$z$ galaxies \citep[e.g.][]{pallottini:17a,pallottini:2017,10.1093/mnras/sty1690,2018MNRAS.474.2884L,2018MNRAS.479..994R}. These studies have identified some important differences between early and local galaxies. Pristine systems tend to have more compact sizes, higher specific SFR, and a more turbulent ISM than their local analogues, consistently with observations \cite[e.g.][]{2009Natur.460..694G,gonzalez2014search,2017Natur.543..397G}.

Moreover, hydrodynamical simulations \citep{behrens2018dusty,2019MNRAS.tmp.2072L,2019MNRAS.487.1844M,2019MNRAS.488.2629A} can be post-processed with physically-rich radiative transfer computations to determine the relation between FIR emission properties and dust mass, temperature, geometry, and chemical composition.
\citet{behrens2018dusty} pointed out that most of the FIR emission in early systems comes from a few massive, UV-opaque, star-forming complexes -- in brief Giant Molecular Clouds (GMC)-- while the UV flux mostly arises from a diffuse ISM component with relatively low optical depth. A spatial segregation of the FIR and UV emitting regions has been indeed observed at $z\sim 6$ \cite[][]{laporte:2017apj,faisst2017high,bowler2018obscured}.  

However, simulations themselves have some limitations due to the demanding computational times that the most detailed ones (in terms of considered physical processes, and resolution) require \citep[][]{Gordon_2001,radtransf_SKIRT,2008Bianchi,Baes_2011,2017MNRAS.470.2791C,behrens2018dusty,2019MNRAS.tmp.2072L,2019MNRAS.487.1844M,bakx20}. In this sense, semi-analytical models, such as the one presented here, are complementary to simulations \citep[][]{F16,popping2017}. The aim of this work is to clarify the physical conditions and emission properties of dust in star forming regions of EoR galaxies. Particularly, we focus on the dust within star forming GMCs, since we expect that the emission from these regions has the strongest impact on the shape of the FIR SED. We refer to \citet{F16} for a complementary treatment including the diffuse ISM.

The paper is organised as follows. In Sec. \ref{GMCs} we derive the main properties of GMCs and compare them with the most updated simulations. We select two prototypical GMCs as representative of the Milky Way (MW) and high-$z$ galaxies population to examine the environmental dependence of their properties. In Sec. \ref{sec4} we compute the effects of star formation on the GMC structure, develop a model for \HII regions, and assess the cloud lifetime against dispersal due to radiative or mechanical processes. Sec. \ref{sec_sub_dust_model} contains the adopted dust model; in Sec. \ref{sub_sec_dust_temperature} we compute the dust temperature in GMCs including the effects of radiation pressure on the dust and gas distribution. In Sec. \ref{IRemiss} the FIR GMC spectra are discussed, and compared with local observations and cosmological simulations. In Sec. \ref{implic} we discuss some important implications of our work; a summary (Sec. \ref{summary}) concludes the paper.

\section{Giant Molecular Clouds}\label{GMCs}

We model GMCs as homogeneous spheres of densities $\rho$ and radius $R$, characterised by a turbulent velocity with a 3D r.m.s. velocity dispersion $\sigma$ and pressure $p$. We assume supersonic turbulence, ($\sigma \gg c_s$, the sound speed of the gas). This simplification is supported by numerical simulation results showing that the ISM of high-$z$ galaxies is highly turbulent \citep{valliniL2018, pallottini:2017}. The density $\rho$ can then be written as $\rho = p/\sigma^2$, neglecting the thermal pressure.

Assuming that GMCs have mass $M$, we introduce the virial parameter
\begin{equation}
    \alpha_{\rm vir}=\frac{5\sigma^2 R}{3 f G M},
\end{equation}
where $f$ is a geometrical factor related to the cloud internal density profile. For spherical clouds with a radial density profile $\rho \propto r^{-\gamma}$ it is $f= (1- \gamma/3)/(1-2\gamma/5)$. We assume $f=1$ for our homogeneous cloud and $\alpha_{\rm vir}=5/3$, which is consistent with local observations \citep{Heyer_2009} and simulations of GMCs \citep{GMCgrisd}. 

\begin{figure*}
	\centering
	\includegraphics[width=0.7\linewidth]{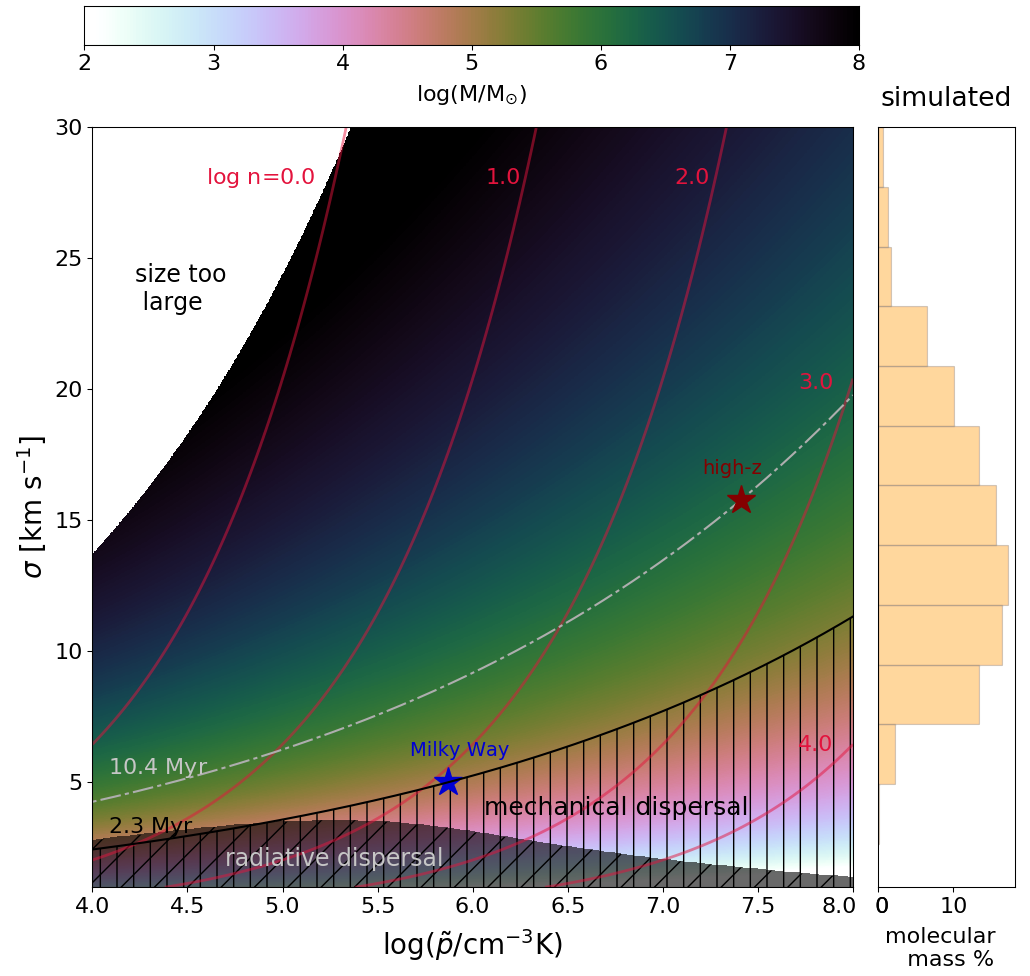}
	\caption{Overview of GMC properties: phase-space diagram of pressure versus dispersion for molecular clouds with masses between $10^{2}\ \mathrm{M_{\odot}}$ and $10^{8}\ \mathrm{M_{\odot}}$. The white region corresponds to clouds that have a size larger than $1$ kpc. Red lines indicate those clouds with number density $\log n = 0.0,\ 1.0,\ 2.0,\ 3.0$, and $4.0$. The grey region shows the clouds dispersed by the expansion of the \HII regions (radiative dispersal) before $4\ \mathrm{Myr}$, while the hatched region indicates those clouds dispersed by either SN explosion or stellar winds (mechanical dispersal) before $2.3\ \mathrm{Myr}$. The blue and brown star marks correspond to a typical GMC in the Milky Way (MW, $\sigma=5\ \mathrm{km s^{-1}}$, $\tilde{p}=10^{5.87}\ \mathrm{cm^{-3}K}$, \citealt{semenov:2018,ddinprep} in prep) and high-z galaxies ($\sigma=15.76\ \mathrm{km s^{-1}}$, $\tilde{p}=10^{7.41}\ \mathrm{cm^{-3}K}$, \citealt{pallottini:2019}), respectively. The grey dashed line represents the mechanical dispersal time for typical high-z GMCs ($t_{\rm d,Myr} = 10.4$) and the black one for the MW-like clouds ($t_{\rm d,Myr} = 2.3$). In the right vertical panel we report the normalised distribution of the $\sigma$ in GMCs found in a zoom-in simulation of a typical high-$z$ galaxy \citep[Freesia, see][]{pallottini:2019}.
	\label{fig1}
	}
\end{figure*}

The cloud mass, radius, and total hydrogen column density ($N_H$) can be written as
\begin{subequations}\label{Jrels}
\begin{align}
    M   &=\frac{1}{2}\frac{\sigma^4}{\sqrt{G^3p}}\\
    R   &=\frac{1}{2}\frac{\sigma^2}{\sqrt{G p}}\\
    N_H &=\frac{1}{2 \mu m_p}\sqrt{\frac{p}{G}}\, ,
\end{align}
\end{subequations}
where $R$ is the Jeans length, $\mu=2$ the mean molecular weight of the gas (we consider only H$_{2}$), and $m_{p}$ the proton mass. The cloud gas number density can be expressed as $n=\rho/ \mu m_{\rm p}$. Relations among the different quantities that enter in our calculations can be visualised in Fig. \ref{fig1}, where we also report the normalised $\sigma$-distribution in GMCs found in zoom-in simulations of a typical, simulated high-$z$ galaxy from \citet{pallottini:2019} and \citet{leung:2019mc}, Freesia. A comparison between the Milky Way (MW) and Freesia is summarised in Tab. \ref{tabgal}.

We note that the r.m.s. dispersion retrieved from this simulation (for further details see \citealt{leung:2019mc}\footnote{Data for $\sigma$ from the simulation are retrieved as a H$_2$ mass weighted PDF on a cell by cell basis. For the first moment of the distribution, there are almost no difference with respect to the extraction of data on a cloud by cloud basis, as shown in \cite{leung:2019mc}.}, and also \citealt{ kohandel:2020}) spans a range $1\lsim \sigma_{\rm kms} \lsim 30$, where $\sigma_{\rm kms}= \sigma/({\rm km \,s}^{-1})$, and has a median value $\sigma_{\rm kms}=15.76$, which is on average higher than what expected from MW-like galaxies at low redshift ($\skms\sim 5$, see observations by e.g. \citealt{larson1981turbulence,rosolowsky2005giant,Heyer_2009} , and simulations by e.g. \citealt{semenov:2018,GMCgrisd}). Derived GMC pressures in high-$z$ clouds are comparable to those found in the Galactic centre, i.e. $ 6 \lsim \log{\tilde{p}} \lsim 8$, where $\tilde p = p/k_B\ \mathrm{cm^{-3}K}$, with $k_B$ being the Boltzmann constant. Simulations \citep[e.g.][]{pallottini:2019}, in agreement with observations \citep{2014ApJ...788...74S}, also find that the effective radii of galaxies at $z=6$ are $< 1$ kpc. In order to exclude unphysically large GMC sizes in our model, we only consider the $p - \sigma$ parameter space satisfying $R<1$ kpc (i.e. we cut off the white region in Fig. \ref{fig1}), corresponding to $\sigma_{\rm kms} < (p/k_B)^{1/4}$. With these assumptions, our model includes GMC masses in the range $2 \le \log (M/\msun) \le 8$. 

\begin{table}
  \begin{center}
    \begin{tabular}{l|c|c|c|c|r}
     \hline
    Galaxy &  $\mathrm{M_{\rm halo}}$ & $\mathrm{M_{*}}$ & $\mathrm{SFR}$ & $z$ & Ref. \\
           &  $ \mathrm{[10^{11}\ M_{\odot}]}$ & $[\mathrm{10^{9}\ M_{\odot}]}$ & $\mathrm{[M_{\odot}/ yr]}$ &    \\
      \hline\hline
      MW & $10^{+3}_{-2}$ & $60.8\pm 11.4$ & $1.65 \pm 0.19$ & 0 &(1)\\
      Freesia & $0.59$ & $4.2$ & $11.5 \pm 1.8$ & 8 &(2)\\
      \hline
    \end{tabular}
    \caption{Properties of the Milky Way and simulated high-$z$ galaxy, Freesia. Data taken from (1) \citet{Licquia_2015,Xue_2008}, and (2) \citet{pallottini:2019}.}
     \label{tabgal} 
  \end{center}
\end{table}

\section{Star formation and HII regions}\label{sec4}

To determine the dust temperature in GMCs we need to estimate the star formation rate (SFR), and the associated flux of UV photons heating the dust grains. In addition, it is important to study the effects of \HII regions both on GMC dispersal and density structure as a result of the dynamical effects occurring during the \HII region evolution. We consider all the stars to be located in the centre of our GMCs.

The SFR can be estimated by assuming a Schmidt-Kennicutt \citep{schmidt1959,kennicutt1998} relation,
\begin{equation}
    {\rm SFR} = \epsilon_{\rm ff} \frac{M}{t_{\rm ff}} = \frac{1}{2}\epsilon_{\rm ff}\frac{\sigma^3}{G}
    = 10^{-5.9} \sigma_{\rm kms}^3 {\msun}{\rm yr^{-1}},
    \label{sfr}
\end{equation}
where we use the relations given in eq. \ref{Jrels} for the cloud mass; $t_{\rm ff} \propto 1/\sqrt{G\rho}$ is the cloud free-fall time, and $\epsilon_{\rm ff}\approx 0.01$ is the amount of gas converted in stars within $t_{\rm ff}$. For the latter value we consider the average computed by \citet{krumholz2007slow}, who analysed a variety of GMCs observed within the MW. Note that in our model the SFR does not depend on pressure, but only on $\sigma$.

In the absence of feedback impairing the star forming ability of the cloud, eq. \ref{sfr} implies a gas depletion time for the GMC equal to $t_{\rm dep} \equiv {M}/{\rm SFR}$, or\footnote{In this paper we adopt the notation $Y_x = Y/10^x$} 
\begin{equation}
    t_{\rm dep,Myr} = \frac{t_{\rm ff}}{\epsilon_{\rm ff}} = 10.5\, \skms \, {\tilde p_8}^{-1/2}\, . 
    \label{tdisp}
\end{equation}
Massive stars produce ionising ($h\nu > 13.6$ eV) photons which create an \HII region around the star forming site. The size of the ionised bubble is set by recombination-ionisation equilibrium, and it is equal to the Str\"{o}mgren radius $R_{\rm S}$:
\begin{equation}\label{RS}
R_{\rm S} =(3 \dot{N}_{\rm i}/4\pi n_{e}^2 \alpha_B)^{1/3}\,,
\end{equation}
which is reached after approximately a recombination time, 
\begin{equation}
t_{r} = (n_{e} \alpha_B)^{-1}.
\end{equation}
In the previous expressions, $\dot{N}_{i}$ is the ionising photon rate, $n_e$ is the electron number density, and $\alpha_B(T=10^4 {\rm K}) = 2.6\times 10^{-13} {\rm cm}^3 {\rm s}^{-1}$ the case-B hydrogen recombination coefficient.
As for gas densities $n_{\rm e} > 1\ \mathrm{cm^{-3}}$, $t_{r} < 2 \times 10^{4}\ \mathrm{yr}$, which is much shorter than the evolutionary timescale of the massive stars driving the \HII region, we can safely assume that the Str\"{o}mgren sphere forms instantaneously. 

We consider that star formation proceeds in a continuous mode, according to a Salpeter single power-law Initial Mass Function (IMF) with stellar mass range $1-100\ \msun$. We assume a metallicity $Z = 0.5\ \mathrm{Z_{\odot}}$. We then use the population synthesis code \code{Starburst99} \citep{leitherer1999starburst99} and the Geneva tracks \citep{1993S..102..339S} to derive the ionising photon rate conversion, $\dot{N}_{\rm i} = \eta_i\ {\rm SFR}/\msun\, {\rm yr}^{-1}$, with $\eta_i = 10^{53.4}$. With these assumptions we find 
\begin{equation}\label{RS1}
R_{\rm S}= 0.14\ \sigma^{7/3}\, {p}\,^{-2/3}= 3.7 \times 10^{-3}\, \skms^{7/3}\, {\tilde p_8}^{-2/3}\,\, \mathrm{pc}.
\end{equation}
Therefore, the \HII region size decreases for increasing GMC pressure, shrinking to (sub-)pc scales when $ {\tilde p_8}\gtrsim 1$ (for $\skms \sim 15$, namely the one of our high-$z$ cloud).

The overpressurised, ionised region begins to expand into the surrounding gas after a sound crossing time, $t_c = R_s/c_i$, where $c_i \approx 10 \, {\rm km\ s}^{-1}$ is the sound speed in the ionised gas. Stated differently, the ionisation front makes a transition from R-type to D-type. It is straightforward to show that the condition $t_r < t_c$ is equivalent to $c_i < n_{\rm e} \alpha_{\rm B} (3\dot{N}_{\rm i}/4 \pi n_{\rm e}^{2} \alpha_{\rm B})^{1/3} \sim 43\ n^{1/3} \sigma_{\rm kms}\, {\rm km\, s}^{-1}$. Hence, we conclude that a R-type phase always precedes the D-type, under realistic conditions.

\subsection{GMC dispersal}\label{tlife}

As \HII regions expand, they inject kinetic energy in the GMC and might ultimately destroy the cloud by dispersing it. In addition to radiation, massive stars produce stellar winds, and ultimately supernova explosions. In the following we provide separate estimates for these effects in terms of their ability to disrupt the cloud and quench star formation. The timescale on which GMC dispersal occurs is crucial to predict the FIR dust luminosity, as will be discussed in Sec. \ref{IRemiss}.

\subsubsection{Dispersal by \HII regions} \label{HII}

The expansion of the \HII region, formed around newly born stars can lead to the dispersal of the cloud. This occurs via two distinct processes: (a) if the Str\"{o}mgren radius exceeds the GMC radius, $R_S > R$, the ionisation front (IF) is density-bounded and the cloud undergoes essentially free-expansion in the lower pressure diffuse ISM, rapidly dispersing. The timescale for the dispersal is approximately equal to the \HII region sound crossing time $t_c=R_S/c_i$; (b) if $R_S < R$, the GMC can still be dispersed during the subsequent D-type expansion of the front, starting at $t_c$.

By using eq. \ref{RS}, we first determine the conditions for which $R_{\rm S}$ exceeds the GMC radius, which is given by eq. \ref{Jrels}. We find that $R_{\rm S}$ is always smaller than the GMC radius for our assumed values of $p$ and $\sigma$, as also reported by recent numerical simulations \citep[in prep.][]{ddinprep}. In fact, the condition $R_{\rm S}>R$:
\begin{equation}\label{RSR}
    \frac{R_{\rm S}}{R} = \frac{0.28 \sigma^{7/3}/p^{2/3}}{\sigma^{2}/\sqrt{G p}} > 1
\end{equation}
implies $n\ll 1\ \mathrm{cm^{-3}}$, corresponding to a cloud radius $>$ kpc for fixed $\sigma$ and $p$, i.e. exceeding the size of a typical galaxy at $z\approx 6$ \citep{2014ApJ...788...74S}. 

Once the IF reaches $R_{S}$ it makes a transition to a D-type front, in which the warm, ionised gas starts to expand, driving a shock into the surrounding neutral gas. To describe this process we follow the approach by \citet{raga2012analytic}. Assuming that (i) the shock is isothermal, (ii) the gas in the ionised region is uniform and in photo-ionisation equilibrium, we can write the velocity of the IF at radius $r$ as:
\begin{equation}\label{Ragavel}
    \frac{1}{c_{\rm i}} \frac{dr}{dt}=\left(\frac{R_{\rm S}}{r}\right)^{3/4}-C\left(\frac{r}{R_{\rm S}}\right)^{3/4},
\end{equation}
where $C=(c_{\rm s}^{2}+\sigma^{2})/(c_{\rm i}^{2}+\sigma^{2})$ and $c_{i},\ c_{\rm s}$ are the sound speeds in ionised and neutral gas, respectively. Note that when $C=0$ the solution by \citet{dyson1980physics} is recovered, and the \HII region expands forever. With the boundary condition $r(t=0)=R_{\rm S}$, eq. \ref{Ragavel} can be integrated analytically to obtain:
\begin{subequations}
\begin{equation}\label{timeR}
    t^{\prime}=\frac{1}{3 C^{7/6}} [q(y)- q(1)]
\end{equation}
with $y=r/R_{\rm S}$, $t^{\prime}=t \sqrt{c_{i}^{2}+\sigma^{2}}/R_{S}$, and
\begin{align}
    q &= -12\  C^{1/6}y^{1/4}+2\sqrt{3}\arctan\left( \frac{\sqrt{3}\ C^{1/6}y^{1/4}}{1-C^{1/3}y^{1/2}} \right)+\nonumber\\
    &+\ln\left[ \frac{(C^{1/3}y^{1/2}+C^{1/6}y^{1/4}+1)(C^{1/6}y^{1/4}+1)^{2}}{(C^{1/3}y^{1/2}-C^{1/6}y^{1/4}+1)(C^{1/6}y^{1/4}-1)^{2}} \right]\,.
\end{align}
\end{subequations}
We now determine the stalling radius ($r_e$) where the \HII region expansion stops. The surviving clouds are those for which $r_{e}<R$. Those not fulfilling this condition are (radiatively) dispersed over the timescale $t^{\prime}(r_{e}/R_{\rm S})$ given by eq. \ref{timeR}. In Fig. \ref{fig1} we identify GMCs with $t<4\ \mathrm{Myr}$ as the black hatched region.
From the above analysis we conclude that \HII regions can only disperse relatively small, $M\le 10^{5}\ \mathrm{M_{\odot}}$, weakly turbulent ($\sigma<5\ \mathrm{km/s}$) clouds. 

\subsubsection{Dispersal by stellar winds and supernovae}

Supernova (SN) explosions and stellar winds can disperse GMCs by accelerating gas to speeds exceeding the cloud escape velocity, $v_e$. The number of supernovae produced by a given SFR at time $t$ can be written as 
\begin{equation}\label{tSN}
    N_{\rm SN}(t) = \Theta(t-t_{\rm SN}) \nu_{\rm SN} {\rm SFR}\, t = 0.02\, \skms^3\, \tmyr \Theta(t-t_{\rm SN}),
\end{equation}
where $\nu_{\rm SN} = (53 \msun)^{-1}$ is the number of SNe per solar mass of stars formed \citep{10.1046/j.1365-8711.2000.03209.x}; the Heaviside function, $\Theta(t-t_{\rm SN})$, accounts for the mass-dependent delay between the onset of star formation and the SN explosion. We consider as a lower limit $t_{\rm SN} \sim 3\ \mathrm{Myr}$, which roughly corresponds to the lifetime of a $100\ \mathrm{M_{\odot}}$ star \citep{Schaerer97}, i.e. the largest mass in our IMF range. The energy released in kinetic form is then
\begin{equation}
    E_{\rm SN}(t) = f_{\rm SN} E_0 N_{\rm SN}(t) =2 \times 10^{48} \skms^3\, \tmyr \Theta(t-t_{\rm SN}) \,\,\, \rm erg. 
\end{equation}
We have assumed a kinetic-to-total energy efficiency $f_{\rm SN}\approx 0.1$ and a standard SN explosion energy of $E_0=10^{51}\, \rm erg$. 

Before the SN explosion, a large amount of kinetic energy is also produced by stellar winds of massive stars,
\begin{equation}
    E_{\rm w}(t) \simeq \frac{1}{3} E_{\rm SN} = 0.67 \times 10^{48} \skms^3\, \tmyr,
\end{equation}
where the ratio of $1/3$ with respect to the SN explosion energy is taken from \code{STARBURST99}. The total mechanical energy is thus $E = E_{\rm SN} + E_{\rm w}$.

Using eq. \ref{Jrels}, the escape velocity of GMCs can be written as a function of $\sigma$ as:
\begin{equation}
    v_e^2 = \frac{2 G M}{R} = 2 \sigma^2 \,.
\end{equation}
For the dispersal condition, we require that the kinetic energy is sufficient to accelerate the GMC gas mass to $v_e$, i.e. $E \ge 0.5\, M v_e^2$. This condition defines a dispersal time $t_d$, at which the cloud becomes unbound, and its gas returned to the diffuse phase. Using the above relations we find
\begin{equation}
    t_{d,\rm Myr} = \frac{t_{d',\rm Myr}}{\Theta(t_{d',\rm Myr}-t_{\rm SN, Myr})+1/3},
    \label{td}
\end{equation}
where $t_{d',\rm Myr}=1.94\times 10^{-3} \skms^3 {\tilde p_8}\,^{-1/2}$ would be the dispersal time assuming that SNe explode at $t_{\rm SN}=0$. Thus, for the expected pressure in high-$z$ galaxies ($\tilde p \simeq 3\times 10^{7}$), GMCs with $\skms \le 15.76$ are dispersed in $\le 10.4\ \mathrm{Myr}$, while in the MW ($\tilde p \simeq 7\times 10^{5}$) typical GMCs with $\skms \le 5$ are dispersed in $\approx 2.3\ \mathrm{Myr}$. 
However, MW clouds can last for $17\ \mathrm{Myr}$ if we consider $\skms \simeq 10 $, i.e. for clouds with $M\sim 10^{6}\ \mathrm{M_{\odot}}$ at the typical MW pressure. Our estimates are consistent with the recent cosmological zoom-in simulations of MW like galaxies by \citet{2019arXiv191105251B}. The authors measure the lifetimes of GMCs with masses above $\ge 10^{5}\ \mathrm{M_{\odot}}$, finding average values below $7\ \mathrm{Myr}$, with less than $1\%$ of clouds living longer than $20\ \mathrm{Myr}$. 

Before dispersal at time $t_d$ (eq. \ref{td}), dust in the GMC efficiently reprocesses the UV light of the stars into FIR emission. When the GMC is finally dispersed, the dust is ejected into the diffuse ISM on much larger scales, thus drastically decreasing its UV optical depth and quenching the FIR emission.

The fraction of GMC gas converted into stars ($f_*= M_*/M$) before dispersal can be derived by combining eqs. \ref{tdisp} and \ref{td}:
\begin{equation}
    f_*= \frac{t_d}{t_{\rm dep}} = \frac{1.8\times 10^{-4} \skms^2}{\Theta(t_{d',\rm Myr}-t_{\rm SN, Myr})+1/3}. 
\end{equation}
Thus, $f_*$ vary from about 1.4\% for a MW-like cloud to 3.2\% for GMCs in high-$z$ galaxies. This is yet another consequence of the higher turbulence level predicted for these systems.
Since the GMC mass fraction converted into stars is low, we assume that the cloud density does not change with time due to star-formation activity. We note that these are lower limits for the amount of stars formed, since we are considering the average free-fall time, neglecting the likely presence of local density inhomogeneities (\quotes{clumps}, \citealt[see e.g.][]{semenov:2018}). These structures may contribute significantly to star formation on very short time scales since $t_{\rm ff}\propto\rho^{-1/2}$. 

The above estimates can be seen as a rough approximation as some unwarranted assumptions have been made. First, eq. \ref{RSR} neglects radiative losses which are likely to occur during the blastwave evolution. Second, off-centre SNe might induce blisters which are not described by our simple geometry. Finally, the stochastic sampling of the IMF might result in values of $\nu_{\rm SN}$ which are different from the average one assumed here.
In spite of these simplifications, eq. \ref{td} provides some useful guidance for the problem at hand, although we leave a more refined treatment to future work. 

\section{Dust model}\label{sec_sub_dust_model}

Following \citet[][herafter \WD01]{weingartner2001dust}, we model dust grains as constituted by a mixture of silicate and carbonaceous grains with a relative mass ratio $\approx 11:1$ \citep[see also][]{Draine03,draine2004astrophysics}. Grains are assumed to be spherical with radii in the range $10^{-3}\ \mathrm{\mu m} < a < 1\ \mathrm{\mu m}$ and with a size distribution given by \WD01:
\begin{subequations}\label{eq_dust_grain_dist}
\begin{align}
\frac{dN^{j}}{da}\ \frac{1}{N_{\rm H}} = \frac{C_{\rm s}}{a}\ \left(\frac{a}{a_{\rm t,s}}\right)^{\alpha_{\rm s}}\ F(a,\beta_{\rm s},a_{\rm t,s})\ G(a,a_{\rm t,s},a_{\rm c,s})\, ,
\end{align} 
where $(C_{\rm s},\ a_{\rm t,s},\ a_{\rm c,s},\ \alpha_{\rm s},\ \beta_{\rm s})$ are five parameters used to fit the observed extinction curves. In particular, we adopt extinction curve appropriate for the MW with $R_{\rm V} = A_{\rm V}/(A_{\rm B}-A_{\rm V})= 3.1$ where $A_{\rm B}$ and $A_{\rm V}$ are the extinctions measured in the {\rm B} ($4400\ \angstrom$) and {\rm V} ($5500\ \angstrom$) spectral bands, respectively. In general, the Small Magellanic Cloud (SMC) is considered a fair local analogue for high-$z$ galaxies because of its low metallicity. Nevertheless, there is no conclusive evidence that high-$z$ dust should be SMC-like \citep[e.g.][]{2010...523A..85G,2011...532A..45S,popping2017,behrens2018dusty,bowler2018obscured}. For this reason, and to allow a direct comparison with local GMCs, we adopt a MW-like dust grain-size distribution and extinction curve. Moreover, we present results only for silicate grains since the contribution of carbonaceous grains is found to be negligible. We neglect the effect of coagulation and grain growth by accretion of gas phase metals occurring in dense environments, which might result in larger grains.

Under these assumptions, the functions $F(a,\beta_{\rm s},a_{\rm t,s})$ and $ G(a,a_{\rm t,s},a_{\rm c,s})$ take the following form:
\begin{equation}
F =\left\{\begin{aligned}
    & 1+\beta_{\rm s}\left(\frac{a}{a_{\rm t}}\right)        \quad & \beta_{\rm s}\ge 0 \\
    & \left[1-\beta_{\rm s}\left(\frac{a}{a_{\rm t}}\right)\right]^{-1} \quad & \beta_{\rm s}<0
    \end{aligned}\right.
\end{equation}
\begin{equation}
G = \left\{\begin{aligned}
    & 1  & 3.5\,\angstrom<a<a_{\rm t,s}\\
    &e^{-[(a-a_{\rm t,s})/a_{\rm c,s}]^3} \quad & a>a_{\rm t,s}
    \end{aligned}\right.
\end{equation}
\end{subequations}

\subsection{Radiation pressure in dusty \HII regions}\label{radpress}

So far we have considered \HII regions only in relation to their eventual dispersal of the GMCs. However, \HII regions can also modify the dust and gas spatial distribution. For instance, radiation pressure from the internal sources, drags both dust and gas outwards (more or less efficiently depending on the strength of the radiation field), resulting in lower central densities.
We now intend to modify our previous assumption of a uniform GMC gas/dust density profile by taking into account this effect. We follow the approach by \citet{2011ApJ...732..100D}, who assumes that the dust-to-gas ratio remains constant throughout evolution. This is equivalent to assuming that dust is well coupled to the gas, so that the radiation pressure acting on the dust directly determines the gas motion. This assumption is valid once that grains approach their terminal velocities (i.e. negligible acceleration). For gas densities $n>10\ cm^{-3}$, and grain sizes in the range $10^{-3}\ \mathrm{\mu m}\le a \le 1\ \mathrm{\mu m}$, the time needed for grains to approach their terminal velocity -- see eq. 28 in  \citet{2011ApJ...732..100D}, is a few tens of years\footnote{More precisely, for the largest (slowest) grains in the two clouds we find $\tau_{\rm drag,MW}= 43\ \mathrm{yr}$, and $\tau_{\rm drag,hz}= 11\ \mathrm{yr}$}. This is much shorter that the timescales of interest here. So we conclude that grains can be safely assumed to be perfectly coupled to the gas.

We can then assume that the gas is in dynamical equilibrium, so that the force per unit volume from radiation pressure is balanced by the pressure gradient:
\begin{subequations}\label{sist_press}
\begin{equation}\label{sist_press_dens}
    n \sigma_{\rm d} \frac{[L_{\rm n}e^{-\tau}+L_{\rm i} \phi(r)]}{4 \pi r^{2} c}+\alpha_{\rm B} n^{2} \frac{\left<h \nu_{\rm i}\right>}{c}-\frac{dp}{dr}=0
\end{equation}
where $\sigma_{\rm d}$ the dust scattering and absorption cross section per H nucleus averaged over the radiation field. $L_{\rm i} \phi(r)$ is the power in ionising photons ($h \nu_{\rm i} \ge 13.6\ \mathrm{eV}$) crossing  a sphere of radius $r$, and $L_{\rm n}$ the one in non-ionising photons (the central cluster UV luminosity is then equal to $L_{\rm n}+L_{\rm i}=L_{\rm UV}$). 
The function $\phi(r)$ and the dust absorption optical depth at the distance $r$ (integrated over wavelength) $\tau(r)$ are determined by the differential equations:
\begin{equation}
    \frac{d \phi}{dr}=-\frac{1}{\dot{N}_{\rm i}}\alpha_{\rm B} n^{2} 4 \pi r^{2}-n \sigma_{\rm d} \phi,
\end{equation}
\begin{equation}
        \frac{d \tau}{dr} = n \sigma_{\rm d},
\end{equation}
\end{subequations}
with boundary conditions $\phi(0)=1$, and $\tau(0)=0$. Following the computation by \citet{2011ApJ...732..100D}, we numerically solve the system eqs. \ref{sist_press}, fixing the pressure at the edge of the \HII region, $p$, and $\dot{N}_{\rm i}$ to the corresponding values for the MW and high-$z$ clouds. 

The resulting\footnote{The above density profiles refer to a steady-state solution of the equations. Hence we have to ensure that the timescale to reach such configuration is shorter than, for instance, the lifetime of massive stars. We compute the grain velocity due to radiation pressure, $v(r)$, by solving the momentum equation, $({L_{\rm UV}}/{4 \pi r^{2} c}) \pi a^{2} Q_{\rm abs} = m_{\rm g} v ({dv/dr})$, where $m_{\rm g}= (4\pi/3) \delta_g a^3$ is the mass of a grain of bulk density is $\delta_g \approx 3\,{\rm g cm}^{-3}$. We estimate the transient timescale as $\Delta t \approx R_{\rm S}/v(R_{\rm S})$, finding $\Delta t \ll 1\ \mathrm{Myr}$. Hence, we can safely assume that the steady-state solution is instantaneously reached.} density profile $n(r)/n$, normalised to our homogeneous Jeans cloud density $n$, is shown in the upper panel of Fig. \ref{fig4} for the high-$z$ GMC. The (normalised) dust number density drops almost to zero for distances $\log x < -1.2$, and reaches $\simeq 3$ at the Str\"{o}mgren radius $R_S$; outside the \HII region the dust profile remains unaffected with respect to the uniform cloud case. 

\begin{figure*}
\includegraphics[width=0.49\linewidth]{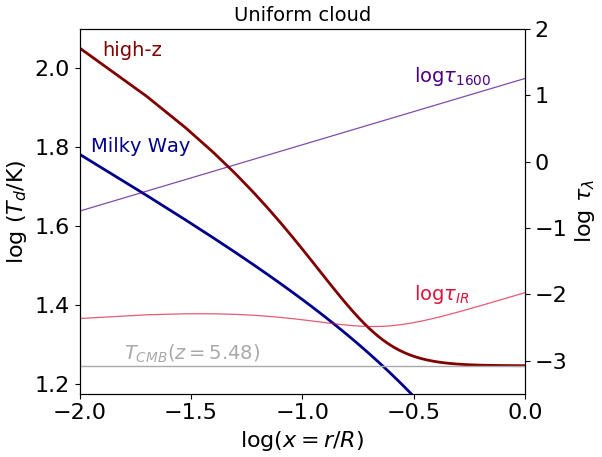}
\includegraphics[width=0.49\linewidth]{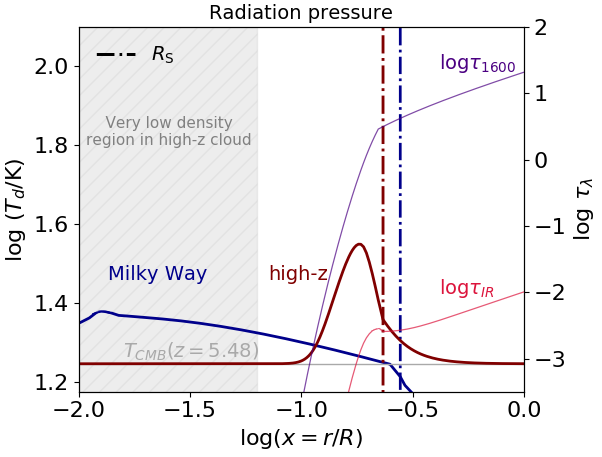}
\caption{Dust temperature vs. normalised distance from the centre of the cloud for typical grains with radius $a=0.1\ \mathrm{\mu m}$.
{\it Left panel}: Case of a uniform GMC in which radiation pressure effects have been neglected. Blue and brown lines show the dust temperature profiles for a MW-like and high-$z$ GMC, respectively. The properties of the two GMCs are the same of those marked as blue and brown stars in Fig. \ref{fig1}, and the stellar age is  $0.4\ \mathrm{Myr}$. The thin lines show respectively the values of the optical depth at $1600 \angstrom$ (purple) impinging on dust grains at various distances from the centre of the cloud, and at the optical depth corresponding to the IR peak wavelength of emission $\lambda_{\rm peak}$ (red, see text) for the high-$z$ GMC.
{\it Right}: as the left panel, but the model takes into account the radiation pressure. The dashed dotted lines represent the location of the boundary of the \HII region in the MW and high-z like cloud.
\label{fig2}
}
\end{figure*}

\section{Dust temperature}\label{sub_sec_dust_temperature}
We now compute the dust temperature, $T_{\rm d}$, by balancing the dust photoheating rate\footnote{We neglect collisional heating.}, $\dot{E}_{+}$, with the emission rate, $\dot{E}_{-}$. To this aim, we consider GMCs to be constituted by a series of concentric shells of dust and gas of radius $r=xR$, with fixed dust-to-gas ratio $D=0.01$. We then compute the temperature of a dust grain located in the shell at radius $x$, $T_{\rm d}(x,a)$, for different values of the grain size in the range $10^{-3}\ \mathrm{\mu m}<a<1\ \mathrm{\mu m}$. 
We adopt the internal radiation field provided by newly formed stars, $F_{\rm int}$, as the only UV radiation source. For the photoheating rate we get 
\begin{subequations}
\begin{equation}\label{tdustshell}
    \dot{E}_{+}= \int_{100\angstrom}^{4000\angstrom} \pi a^{2} Q_{\rm abs}(\lambda,a) (1-e^{-\tau_{\rm shell} }) F_{\rm int} {\rm d}\lambda\, ,
\end{equation}
\begin{equation}\label{e+int}
F_{\rm int} = \frac{L_{\rm \lambda}}{\pi (xR)^{2}} e^{-\tau_{\lambda}}\,.
\end{equation}
\end{subequations}
The integral runs over the range $ 100\ \angstrom< \lambda < 4000\ \angstrom$ since the UV flux out of this range contributes negligibly to grain heating in actively star-forming systems \citep{buat1996star}. $L_{\rm \lambda}$ is the specific luminosity of the central cluster; using the SFR of our clouds (eq. \ref{sfr}) we can write $L_{1500 \angstrom}= \eta_{1500\angstrom}$ SFR and compute $\eta_{1500\angstrom}$ using \code{STARBURST99} (see Sec. \ref{sec4} for the detailed assumptions). We find that $\eta_{1500\angstrom}$ increases with the cloud lifetime, ranging from $10^{41.8}\ \mathrm{erg\ s^{-1}/\msun\ yr^{-1}}$ at $0.4\ \mathrm{Myr}$ of the cloud age, to $10^{43.1}\ \mathrm{erg\ s^{-1}/\msun\ yr^{-1}}$ at $10\ \mathrm{Myr}$.
The dust optical depth can be written as:
\begin{equation}\label{radtau}
    \tau_{\lambda} = \left[\tau_\lambda\right]_0^{R} =\int_{0}^{R} n(r) dr \int_{a_{\rm min}}^{a_{\rm max}} \pi a^{2}\ Q_{\rm abs}(\lambda,a)\ \frac{dN}{da}(a)\ {\rm d}a \,.
\end{equation}
Within a shell at radius $r$, and width $2dr$, the optical depth is $\tau_{\rm shell} = \left[\tau_{\lambda}\right]_{r-dr}^{r+dr}$. $dN/da$ is the grain size distribution given in eq. \ref{eq_dust_grain_dist}, and $n(r)$ is the number density of the cloud, which is either uniform, or is computed via eq. \ref{sist_press_dens} when taking into account the effect of radiation pressure. 

The grain radiates energy at a rate given by: 
\begin{align}\label{eq_energy_loss_rate}
\dot{E_{-}} = 4\pi \int_{0}^{\infty}  \frac{4}{3}\pi a^{3} \delta_g B_{\lambda}(T_{\rm d}) k_{\lambda} d\lambda\,,
\end{align}
where $B_{\lambda}$ is the Planck function and $k_{\lambda}(\lambda)$ the opacity per unit mass, which can be well approximated by a power law (for $\lambda > 20\, \mum$) as in \citet{draine2004astrophysics}:
\begin{align}\label{kabs}
k_{\lambda}(\lambda) = k_{\rm abs}\ \left(\frac{\lambda}{\mathrm{cm}}\right)^{-\beta_{\rm d}}
\end{align} 
the constant of proportionality in the MW extinction curve corresponds to $k_{abs}=2.47\times 10^{-3}\ \mathrm{cm^{2}/g}$. Estimates of the power slope found in literature span the range $1 \le \beta_{\rm d} \le 2.5$ \citep[e.g.][]{1983QJRAS..24..267H,obs_betad,2003...404L..11D,2008...481..411D,2009...506..745P,2010...520L...8P,2011...536A..25P,2011...536A..19P}; here we adopt $\beta_{\rm d}=2$ as in \citet{draine2004astrophysics}. 
The equilibrium temperature\footnote{ Very small grains might undergo stochastic temperature fluctuations \citep[e.g.][]{1989stocheat,stocheat_Draine}. We neglect this effect in the paper, although we recognise that it might influence the observed emission under certain conditions. } $T_{\rm d} (a, \beta, \Sigma_{\lambda})$ can be then obtained by imposing $\dot{E}_{+} = \dot{E}_{-}$. Furthermore, we take into account the effect of the Cosmic Microwave Background (CMB) acting as a thermal bath. We always correct the dust temperature \citep{da2013effect}:
\begin{align}\label{eq_dachuna_effect}
T_{\rm d}' = \{T_{\rm d}^{4+\beta_{\rm d}}+T_{0}^{4+\beta_{\rm d}}[(1+z)^{4+\beta_{\rm d}}-1]\}^{1/(4+\beta_{\rm d})},
\end{align}
where $T_{0}=2.73\ \mathrm{K}$ is the CMB temperature at $z=0$. 

In Fig. \ref{fig2} we show the results for $T_{d}$ with/without the effects of radiation pressure on the dust density, for the typical grain radius $a = 0.1\ \mathrm{\mu m}$. We compare dust temperatures for the prototypical MW and high-$z$ GMC (blue and brown star in Fig. \ref{fig1}, respectively) at $0.4\ \mathrm{Myr}$ of stellar age, i.e. well before stellar feedback effects start to play a role. In the following, we analyse separately the results for uniform clouds, and the case including radiation pressure due to UV radiation. 

\begin{figure*}
    \centering
    \includegraphics[width=0.8\linewidth]{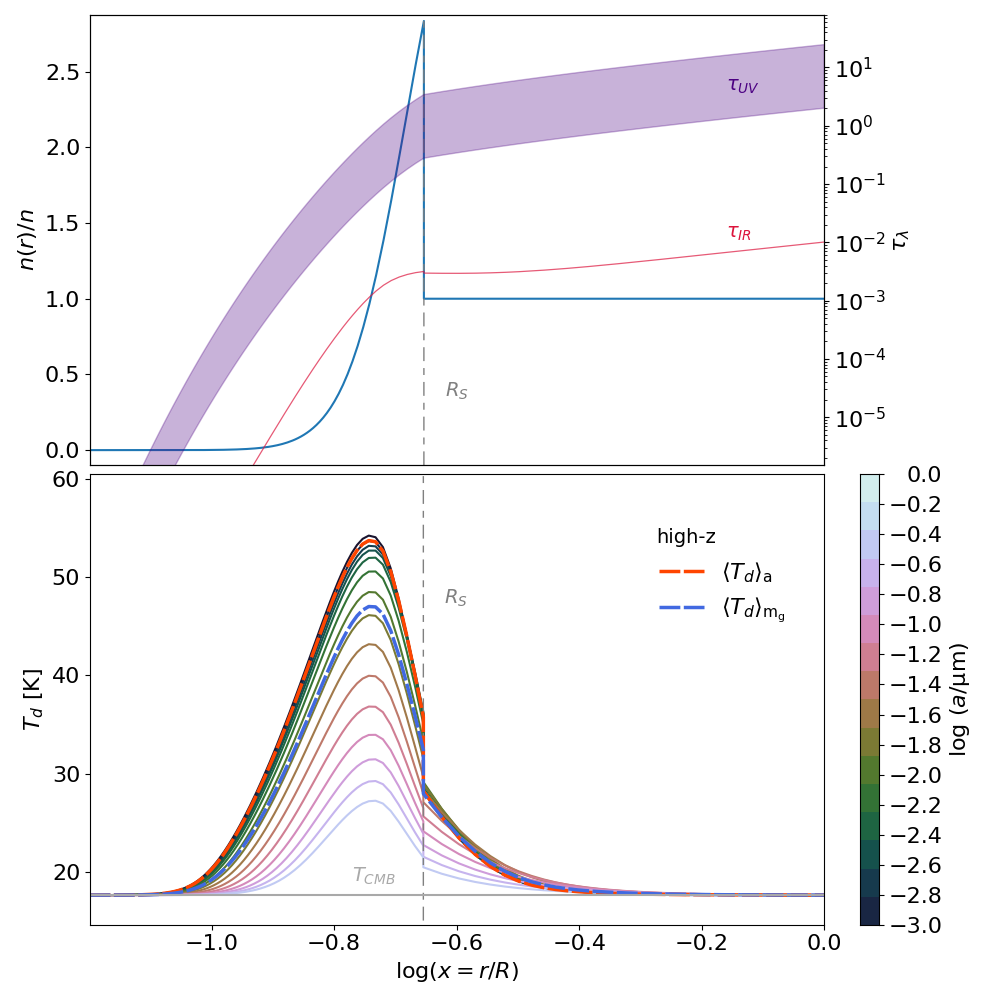}
    \caption{{\it Upper panel}: Dust density modified by the radiation pressure in the high-$z$ GMC as a function of the normalised distance $x$ from the centre. We also show the dust optical depth to UV radiation in the wavelength range $1500\ \angstrom<\lambda<3100\ \angstrom$ (purple shaded region, computed via eq. \ref{radtau}), along with the dust IR optical depth (red solid line) at the peak emission wavelength. {\it Bottom}: Grain temperature for the high-$z$ GMC as a function of radial distance from the centre and for different grain radii, colour-coded in the colourbar. The dashed line shows the size-averaged temperature, $\Tda$. We consider a cluster stellar age $0.4\ \mathrm{Myr}$.
    \label{fig4}
    }
\end{figure*}

\subsection{Uniform cloud}\label{unifcloud}

In the left panel of \ref{fig2}, we show that in the central region of uniform clouds ($\log x<-1$), dust is warmer by $\Delta T_{\rm d}\sim 50\ \mathrm{K}$ in high-$z$ GMCs with respect to MW ones. The higher pressure of high-$z$ GMCs results in larger $N_H$ and $\tau$ values; radiation is more efficiently absorbed in the vicinity of the source, and therefore dust becomes warmer\footnote{Although the higher SFR in the high-$z$ cloud results in a larger $F_{\rm int}$, this does not imply a higher $T_d$. We show below (eq. \ref{Tdp}) that the dust temperature in the GMC is virtually independent of the SFR.}. To show this more quantitatively, we have plotted in Fig. \ref{fig2} the optical depth $\tau_{1600 \angstrom}$ as a function of $x$. The dust mass heated by the internal source is essentially that contained within a radius, $r_d$, such that $\tau_{1600 \angstrom}(r_d)=1$. For clouds with higher $p$, $r_d$ decreases accordingly because of the higher column density (see eq. \ref{Jrels}), resulting in higher dust temperatures. This finding provides the physical basis to interpret the simulation results by \citet{behrens2018dusty}. The authors, in fact, noticed that the galaxy regions with the highest dust temperature, which contribute to most of the FIR radiation, are optically thick. 

Differences between early and local GMCs are also seen in the external layers ($\log x \gtrsim -0.5$). Due to the absorption of the UV radiation within the cloud, and in the absence of an external radiation field (such as the interstellar radiation field), $T_{d}$ drops at the surface of the cloud down to the CMB temperature. The CMB acts as a thermal bath, setting a temperature floor for dust around $20\ \mathrm{K}$ at $z\gsim 6$ (see eq. \ref{eq_dachuna_effect}). This effect is sub-dominant in the central regions of the GMC, where dust reaches higher temperatures $T_{\rm d} \gg T_{\rm CMB}$. In Fig. \ref{fig2} we computed the CMB correction for $z = 5.48$, i.e. the mean redshift of the \citet{capak15} sample. 

We can understand these results in simple physical terms. Let us assume that the amount of dust heated by the internal radiation field is that contained within $r_d$, the distance at which the dust optical depth is $\tau_{1600 \angstrom} \approx 1$. For a gas of solar metallicity ($Z=Z_\odot$) and dust-to-gas ratio ${\cal D}={\cal D_\odot}=0.01$, the condition $\tau_{1600 \angstrom} =1$ is reached for a column density $N_1 = 1.4\times 10^{21} {\rm cm}^{-2}$. By using eq. \ref{Jrels}, we find that the dust content within $r_d$ is
\begin{equation}\label{Mdp}
    M_d(r_d) =  {\cal D} (\mu m_p)^3 N_1^3\frac{\sigma^4}{p^2}. 
\end{equation}

As the UV light is completely absorbed by dust within $r_d$ and re-emitted FIR wavelengths, we can write $L_{\rm FIR}\sim L_{\rm 1500\angstrom}= 10^{2.9} \sigma^3_{\rm kms } \mathrm{L_\odot}$. We can finally derive the dust temperature adapting to our dust model the formula from \citet{10.1111/j.1365-2966.2009.16164.x}, 
\begin{equation}\label{Tdp}
    T_d = 3.24 \left(\frac{L_{\rm FIR}/\lsun}{M_d/\msun}\right)^{1/6}\, {\rm K} = 122\ {\tilde p_8}^{1/3} \sigma^{-1/6}_{\rm kms}\, {\rm K}.
\end{equation}
This result shows that star forming GMCs at higher pressure contain hotter dust. For example, for a pressure $p/k_B = 10^{7.41}\, \pk$, and $\sigma_{\rm kms}=15.76$ (typical of high-$z$ simulated galaxies, \citealt{pallottini:2019}), we find $T_d\sim 50\ \mathrm{K}$, in good agreement with the numerical result in Fig. \ref{fig2}. Although velocity dispersion modifies the dust temperature as well, its effect is less relevant than pressure. As already mentioned, $T_d$ is virtually independent on SFR: as $\sigma \propto \mathrm{SFR}^{1/3}$, from eq. \ref{Tdp} we deduce $T_{\rm d}\propto \mathrm{SFR}^{-1/18}$. Hence pressure is the primary factor controlling dust temperature. Thus, the physical explanation for the higher $T_d$ in highly pressure environments is related to the limited amount of dust required to block the UV radiation. 
Dust located outside $r_d$ cools to the CMB temperature as the flux from the stars is largely blocked.

So far we have assumed that the GMC is optically thin with respect to the FIR radiation. We motivate such choice by showing in Fig. \ref{fig2} the optical depth $\tau_{\rm IR}$ at the peak wavelength of emission, computed from Wien's law: $\lambda_{\rm peak} T_{d}(x)= 0.29\ \mathrm{cm\ K}$. We find that $\tau_{\rm IR}\ll 1$, i.e. dust is optically thin with respect to its FIR emission, even at this very high densities. Indeed, we can understand this by considering the optical depth at $100\ \mathrm{\mu m}$ from \citet{Draine03}: $\tau_{100\ \mathrm{\mu m}}/N_{H}=5.07 \times 10^{-25}$, hence\footnote{As a reference, the column density at the surface of our typical high-$z$ cloud is $N_{H}=3.7\ \times 10^{22}\ \mathrm{cm^{-2}} $.} in order to have $\tau_{100 \mathrm{\mu m}}=1$, $N_{H}=1.9 \times 10^{24}\ \mathrm{cm^{-2}}$.
Using eq. \ref{Jrels}, we can compute the pressure needed to reach such value in our GMCs and we find $\tilde{p}_{8}=64$ which is outside the considered range.  If these high pressures were to be reached, dust might become optically thick to FIR emission as well \cite[see e.g.][]{Conley_2011,Cunha_2015}.

\subsection{Radiation pressure-modified density profile}\label{tdradpress}

Next, we consider dust temperature variations associated with the density redistribution induced by radiation pressure on dust described in Sec. \ref{radpress}. The obtained profiles of the dust temperature as a function of the distance are shown in the right panel of Fig. \ref{fig2}. By comparing with the uniform cloud (left panel), we see that the inclusion of radiation pressure decreases $T_d$: for the high-$z$ (MW) cloud $T_d$ spans the range $17-40$ K ($2.73-25$ K). 
The lower $T_d$ in both cases is associated to the fact that dust is pushed at larger distances from the UV source where the flux is more geometrically diluted. We note that the resulting central region devoid of dust is much wider in the high-$z$ GMC, where it extends up to $\log x=-1.2$ (see the grey hatched region). In the MW cloud this region is much smaller, $\log x \sim -3$.
This is due to the larger UV luminosity resulting from the higher SFR, in turn associated with larger dispersion in the high-$z$ GMC (SFR$\propto \sigma^3$, see eq. \ref{sfr}). In both cases (high-$z$ and MW clouds) the external layers are unaffected by the presence of the internal sources since the emitted radiation has already been absorbed within the cloud. Hence, no appreciable differences with respect to the uniform density case are found at large distances (in the absence of the external interstellar radiation field).

The dust temperature depends also on grain size, $a$, as shown in Fig. \ref{fig4} in the case of the high-$z$ cloud. We can isolate the terms related to the grain size in the photoheating and emission rates (eq. \ref{eq_energy_loss_rate} and \ref{e+int}), and find that $T_d\propto [Q_{abs}(\lambda,a)/a]^{1/6}$.
The dependence of $T_d$ on wavelength and grain radius is non-monotonic. Blueward of $\lambda=1500\ \angstrom$ smaller grains are more efficiently heated; at longer wavelengths $2000\ \angstrom<\lambda<3500\ \angstrom$, intermediate grains with $a\sim 0.1\ \mathrm{\mu m}$ have the highest $Q_{\rm abs}$ value. This has a noticeable effect on the dust grains temperatures at different radii in the GMC. 

While within the \HII region the photo-heating rate is dominated by short-wavelength UV radiation with $\lambda <2000\ \angstrom$, radiation with longer wavelengths ($\lambda \gtrsim 3000\ \angstrom$) takes over in the external neutral layers. As a result, near the centre of the cloud ($\log x< -0.6$), the small grains, $a\sim 10^{-3}\ \mathrm{\mu m}$, are the hottest, reaching $T_{d}>50\ \mathrm{K}$. At the GMC surface, instead, intermediate size grains, $a\sim 0.1\ \mathrm{\mu m}$, attain the highest temperatures. These trends are clearly visible in the bottom panel of Fig. \ref{fig4}, where we also show the size-averaged temperature $\Tda$ and grain mass-averaged temperature $\langle T_{\rm d}\rangle_{\rm m_{\rm g}}$ (averaged over $dN/da \times 4 \pi a^{2} da$). At the surface of the cloud, independently of the grain size, the dust temperature converges to the CMB temperature due to the very high optical depth (see the purple shaded region in fig. \ref{fig4} for the dependence of the grain size-averaged UV optical depth on the distance from the centre of the cloud). 

We pause here for a remark. We considered the same continuous SFR described in Sec. \ref{sec4} also in the radiation pressure modified density profile treatment. This is clearly an approximation, since the reduction in the central density of the cloud (shown in the upper panel of Fig. \ref{fig4}) actually affects the SFR, which is proportional to gas density. However, star formation is distributed all over the cloud, and not just in its centre (as shown also by recent fully resolved simulations, see \citealt{ddinprep} in prep.). Therefore, the average SFR over the whole cloud will not necessarily change considerably, despite the local density drop. Although this approximation over the SFR has been introduced, we prefer to include radiation pressure in our models, since we have shown that it alters considerably our predictions on dust temperatures. Moreover, we underline that our main goal is to give a lower limit for the dust temperature in GMCs, and to show how this lower bound evolves from low to high-$z$. The radiation pressure case gives a lower temperature value with respect to the homogeneous case, and yet a higher $T_{\rm d}$ with respect to what so far assumed at high-$z$.

\begin{figure*}
    \centering
    \includegraphics[width=0.67\linewidth]{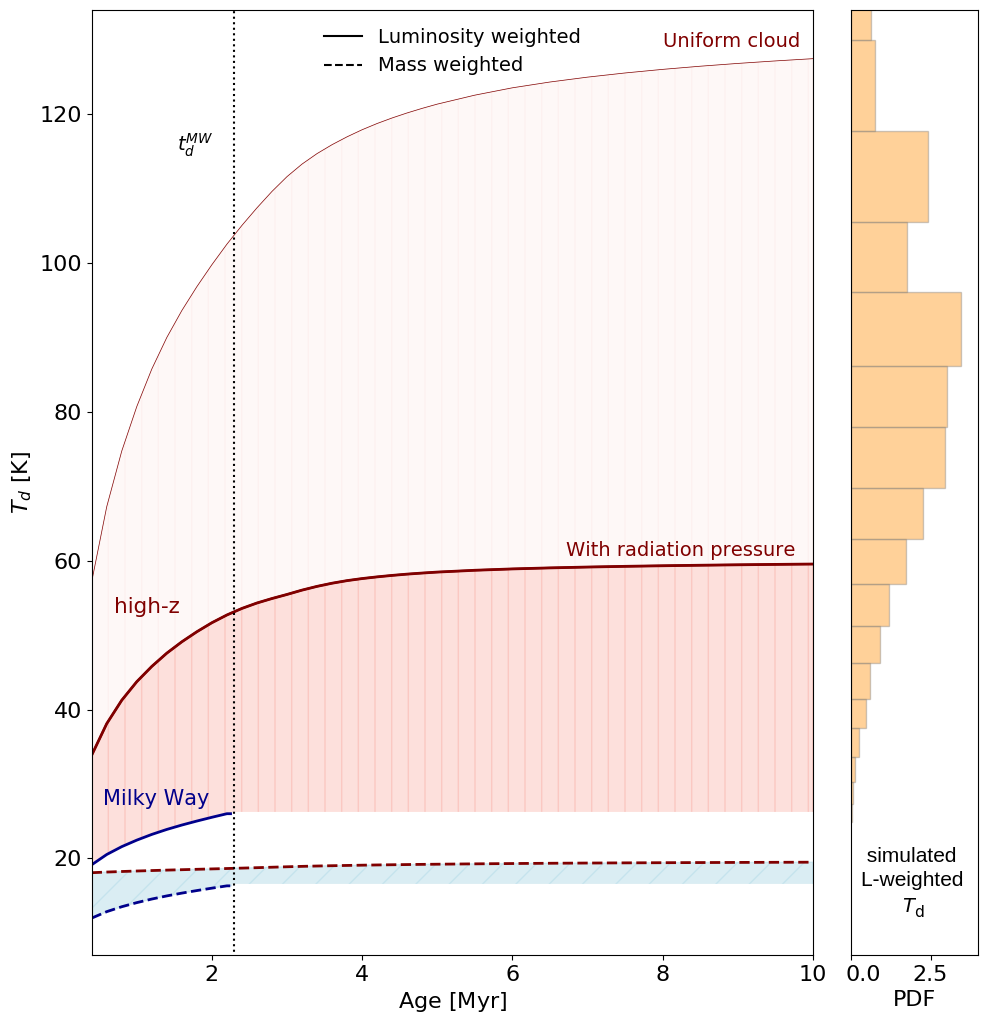}
    \caption{Mass weighted $\TdM$ (dashed line) and luminosity weighted $\TdL$ (solid line) dust temperatures at various ages of the cluster and hosting GMC (see Sec. \ref{masswelumwe}). The blue line corresponds to the MW cloud, while the brown one to the high-$z$ cloud, already described in Fig. \ref{fig1}. For the thick lines we always consider the effect of radiation pressure, and we account for the stellar emission evolution until the dispersal of the clouds, which happens at different times in the two cases (see Sec. \ref{tlife}). The thin brown line corresponds to the uniform high-$z$ cloud case, instead. The grey dotted line represents the dispersal time of the MW cloud. The vertical panel shows the normalised distribution of the luminosity weighted dust temperatures in the simulations by \citet{behrens2018dusty}.
    \label{fig6}
    }
\end{figure*}

\subsubsection{Mass- and luminosity- weighted dust temperature}\label{masswelumwe}

From the above results we can compute the mass-weighted, $\langle T_{\rm d}\rangle_M$, and luminosity-weighted, $\langle T_{\rm d}\rangle_L$, dust temperature as a function the age of the stellar cluster accounting for the associated evolution of the UV luminosity (taken from \code{STARBUST99} with the usual assumptions on the SFR and metallicity, see Sec. \ref{sec4}). We compute $T_d$ in the time interval $0-t_{\rm d,Myr}$, where $t_{d,Myr}$ is the GMC lifetime against dispersal caused by feedback processes (Sec. \ref{tlife}). As usual, in Fig. \ref{fig6} we compare high-$z$ and MW GMCs.

For the high-$z$ cloud, most of the dust mass is in thermal equilibrium with the CMB, $\TdM^{{\rm hi}-z}\sim T_{\rm CMB}$. However, a small fraction of the dust located near the UV source is much hotter, and contributes strongly to the luminosity-weighted temperature. Indeed, we find $\TdL^{{\rm hi}-z}\sim 60\ \mathrm{K} \gg \TdM^{{\rm hi}-z}$ at $10\ \mathrm{Myr}$ of the central cluster stellar age. The luminosity-weighted temperature is of particular importance as it represents a good proxy for the dust temperature that would be inferred from the SED fitting of the source. Neglecting the effects of radiation pressure, would lead to even higher temperatures, $\TdL^{{\rm hi}-z}\sim 100\ \mathrm{K}$. This is because more dust would be found in the proximity of the source (at $\log x< -1$), as already discussed (see Sec. \ref{tdradpress}). The MW GMC is instead much colder, with $\TdM^{\rm MW}\le 15\ \mathrm{K}$ and $\TdL^{\rm MW} < 30\ \mathrm{K}$. This is because at $z=0$ (i) the CMB temperature sets a much lower floor for $T_{d}$, and (ii) the more diffuse structure of the cloud results in a less efficient dust heating (see Sec. \ref{unifcloud}). We will return to this point in the following Section. 
\section{Infrared emission}\label{IRemiss}

\begin{figure}
    \centering
    \includegraphics[width=1.0\linewidth]{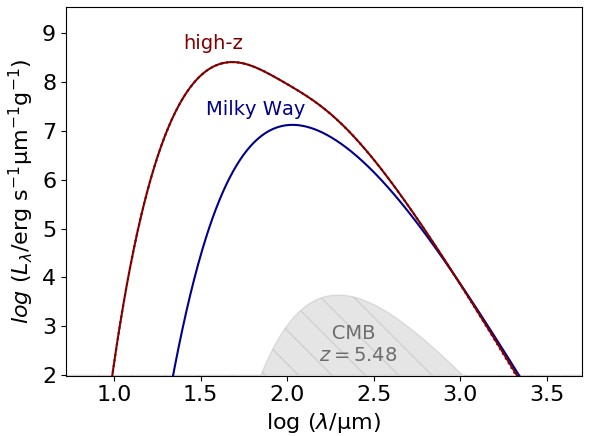}
    \caption{Specific luminosity per unit wavelength of the high-$z$ cloud (brown line) and MW cloud (blue line) at the stellar age of $0.4\ \mathrm{Myr}$. We also show as a comparison the CMB specific luminosity at redshift $z=5.48$ (grey line). The latter is important because at such redshift, dust emission is observed against it, hence the intensity of the CMB must be subtracted out.
    \label{fig7}
    }
\end{figure}

We predict the infrared luminosity of our star-forming GMCs by using the above grain mass-averaged temperature $\langle T_{\rm d}\rangle_{\rm m_{\rm g}}$ (blue dashed line in Fig. \ref{fig4}), and weighting the contribution of each shell by its dust mass, $m_{\rm d, shell}$:
\begin{equation}\label{lumIR}
    L^{i}_{\lambda}=\frac{8 \pi h c^2}{\lambda^5}\frac{k_\lambda}{M_{\rm d}}  \int_{0}^{1} \frac{m_{\rm d, shell} (x)} {\exp(hc/k_{\rm B}\langle T_{\rm d}\rangle_{\rm m_{\rm g}}\lambda)-1} dx. 
\end{equation}
The comparison between the high-$z$ and MW cloud is shown in Fig. \ref{fig7} at the stellar age of $0.4\ \mathrm{Myr}$. As expected, the IR emission from the high-$z$ cloud is $\sim 30$ times stronger due to its higher star formation rate, entailing a stronger UV field within the cloud. In fact, the SFR in the two GMCs differs by a factor $\approx 30 \times$ ($\mathrm{SFR}^{{\rm hi}-z}=4.8 \times 10^{-3}\ \mathrm{\msun yr^{-1}}$ vs. $\mathrm{SFR}^{\rm MW}=1.5 \times 10^{-4}\ \mathrm{\msun yr^{-1}}$). Note that the different $T_d$ shifts the emission peak from $100\ \mathrm{\mu m}$ (MW) to $\sim 47\ \mathrm{\mu m}$ at high-$z$. 

Moreover, we fit the above IR spectra to compare the recovered temperature with the luminosity-weighted dust temperature from the model. As discussed in the Introduction, the two should correspond. Nevertheless, this sanity check is important since dust temperature in our model has a distribution depending on the grain size and position within the GMC. In the fitting procedure, we use a single temperature grey body function $B_{\lambda}(T_{\rm d})k_{\lambda}$, with $k_{\lambda}$ as in eq. \ref{kabs}. We leave $\beta_{\rm d}$ and $T_{d}$ as free parameters (in the range $1.5<\beta_{\rm d}<2.5$, and $10\ \mathrm{K}\le T_{\rm d}\le 70\ \mathrm{K}$). The recovered temperatures are consistent with $\TdL$ for both the MW and high-$z$ cloud: 
\begin{subequations}
\begin{equation}
    T_{\rm d,fit}^{\rm MW}= 19 \pm 1\ \mathrm{K}\\
    T_{\rm d,fit}^{{\rm hi}-z}= 40 \pm 2\ \mathrm{K}
\end{equation}
where the corresponding values for the luminosity weighted temperatures are:
\begin{equation}
    \TdL^{\rm MW}= 19\ \mathrm{K}\\
    \TdL^{{\rm hi}-z}= 37\ \mathrm{K}
\end{equation}
\end{subequations}
We have repeated the fitting procedure also for the high-$z$ cloud at $10\ \mathrm{Myr}$ of stellar age: we find $T_{\rm d,fit}^{{\rm hi}-z}= 62 \pm 3\ \mathrm{K}$, consistent with our $\TdL^{{\rm hi}-z}= 60\ \mathrm{K}$. This is particularly interesting and useful, as $T_{\rm d,fit}$ represents the temperature generally used for extrapolating IR luminosity.

Another quantity of interest that can be deduced from our model is the FIR luminosity per unit mass of gas in the GMC.
We integrate the FIR luminosity over wavelength and divide by the mass of the cloud, and for the high-$z$ GMC at the end of its lifetime ($10.4\ \mathrm{Myr}$) we find:
\begin{subequations}
\begin{equation}\label{LMhz}
\frac{L_{\rm FIR}}{M} \sim  10\ \mathrm{L_{\odot}/ M_{\odot}}\, .
\end{equation}
This value exceeds by $\sim 10 \times$ the one we find in our MW cloud, where we have at the end of its lifetime ($2.3\ \mathrm{Myr}$): 
\begin{equation}
L_{\rm FIR}/M \sim   1\ \mathrm{L_{\odot}/M_{\odot}}\, .
\end{equation}
\end{subequations}
This value is consistent with the average value observed in MW GMCs in which the \HII region is obscured \citep{1989ApJ...339..149S}. 

As a caveat, we warn that so far we have discussed the intrinsic luminosity of our clouds. In order to compute the observed luminosity, one would need to account for the fact that dust emission is observed against the CMB; hence, the latter must be subtracted out \cite[as in][]{F16}. However, as $\TdL \gg T_{\rm CMB}$ in our clouds, such correction is negligible, as it can be realised from an inspection of Fig. \ref{fig7} where the dust and CMB spectra are compared. 

Our results show a colder mass-weighted, but consistent luminosity-weighted dust temperature with respect to simulations by \citet{2019MNRAS.tmp.2072L} (see Sec. \ref{intro}). We also confirm that mass-weighted dust temperatures are in general significantly lower than luminosity-weighted ones. Indeed, at redshift $z=6$ they find that the bulk of the dust mass is as cold as $\TdM=30.7\ \mathrm{K}$, while from SED fitting they find the equivalent temperature to be around $T_{\rm d,eqv}\sim 45-50\ \mathrm{K}$. These modest quantitative differences are not surprising as in this work they are considering very compact systems, optically thick to IR, which is not the case of our clouds (see discussion at the end of Sec. \ref{unifcloud}). Moreover our GMCs are not exposed to an external radiation field, thus approaching the surface dust temperature reaches the CMB temperature (and also $\TdM \sim T_{\rm CMB}$). Nevertheless, this sort of bi-modality in the dust temperature (see description in Sec. \ref{intro}) is in accord with the results of our model (see Fig. \ref{fig6}), and we find a physical motivation for this behaviour as extensively described in Sec. \ref{masswelumwe}. 

The dust temperature distribution that we infer within our high-$z$ GMC, is also consistent with the findings by \cite{2019MNRAS.488.2629A}. At redshift $z\sim 6$, they find that in the outer low-density regions of their simulated galaxies, a large amount of dust is in thermal equilibrium with the CMB. Instead, for more centrally-concentrated dust the temperature increases up to $\sim 100\ \mathrm{K}$ due to a stronger UV flux. 

We conclude this Section by comparing our results also with the simulations by \citet{behrens2018dusty}, where the average $\TdL=91\pm23\ \mathrm{K}$ at $z=8.38$ \citep[c.f.r.][]{laporte:2017apj}. We underline that, even at the remarkably high resolution ($\approx 30$ pc) of their simulations, dust temperature effects related to the radiation pressure in \HII regions included here, cannot yet be properly treated (see also discussion in \citealt{pallottini:2019,decataldo:2019}). This explains the hotter dust temperatures found in their simulations, which lie in between the two curves corresponding to the two models (homogeneous and with a density profile due to radiation pressure, see vertical panel in Fig. \ref{fig6}).

In conclusion, all our results indicate that the dust temperature increases with redshift. This is confirmed also by recent observations \citep[see e.g.][]{2019MNRAS.487L..81L,bakx20}. This evidence is not accounted in most observations, which usually assume $T_{\rm d} = 35-45\ \mathrm{K}$ at $z>5$. This conclusion bears important implications that we are going to discuss next.

\section{Implications}\label{implic}
The conclusion reached so far, namely that dust is on average warmer at early cosmic times, entails a number of important implications that we briefly discuss here. 

\subsection{Obscured UV emission}\label{galfut}

At redshift $5\le z \le 10$ the Star Formation Rate Density (SFRD) is usually probed in the rest-frame UV \citep[e.g.][]{oesch2012bright,Bouwens16,finkelstein2012candels}, but at the moment there is no consensus on the obscured fraction of UV emission at these early epochs. Clearly, not taking into account obscuration, can lead to underestimate the star formation rate in early structures. Interestingly, we find that UV photons emitted by newly-born stars are completely absorbed within our high-$z$ GMCs up to the time of dispersal, $t_d$, when the gas is removed by stellar feedback. This feature is not appreciably modified by the inclusion of radiation pressure and the subsequent formation of a low-density cavity in the central regions of the GMC. Hence, the unobscured UV emission observed in high-$z$ galaxies should come from clouds that have already been dispersed (see sec. \ref{tlife}), either due to mechanical or radiative processes (Sec. \ref{tdisp}). In short, the obscured fraction depends on cloud lifetimes. 

To get a rough estimate of the fraction of obscured UV emission within our typical high-$z$ GMC, we integrate the UV luminosity at $1500\ \angstrom$ for continuous star formation until $t_d$\footnote{In practice we approximate the continuous SFR with a series of subsequent bursts, and the resulting $f_{\rm UV, obscured}$ converges considering more than $20$ bursts.}. After the cloud is dispersed, all the UV photons leaks away. We can then compute the ratio between obscured UV photons (those produced before $t_d$) and all the produced UV photons:
\begin{equation}
    f_{\rm UV, obscured}=\frac{\int_{0}^{t_{\rm d, Myr}} L_{1500 \angstrom}(t)\ dt}{\int_{0}^{100 Myr} L_{1500 \angstrom}(t)\ dt} 
\end{equation}
where the upper limit of $100\ \mathrm{Myr}$ is indicative of the time at which the UV emission from the past star formation episodes becomes negligible. We find that $f_{\rm UV, obscured} \simeq 0.42$, i.e. almost half of the energy in the UV produced by star formation in our high-$z$ GMC is obscured. Note that in our MW cloud this fraction is much lower, around $f_{\rm UV, obscured} \simeq 0.18$, basically due to the shorter lifetime of the cloud. Our finding is consistent with recent high-$z$ observations. 

So far, dust obscuration has been observed at $z\sim 7$ by \citet{bowler2018obscured} in six bright LBGs. They were able to obtain a $5\sigma$ detection of the highest redshift galaxy targeted, but no detection of the five remaining sources. From a stacking analysis they determined the average FIR luminosity of the sample to be $L_{\rm FIR} \sim 2\times 10^{11}\ \mathrm{L_{\odot}}$. Then, converting this observed FIR luminosity into a star-formation rate, they found that $\approx 50\%$ of the total SFR is obscured by dust. Moreover, \cite{2020arXiv200410760F} recently observed that within their UV-selected sample, massive galaxies ($\log (M_{*}/M_{\odot})>10)$ at $z\sim 5-6$ exhibit an obscured fraction of star formation of $\sim 45\%$. All together these results indicate a rapid build-up of dust during the epoch of Reionization.

\subsection{IRX-$\beta$ relation}

 In Sec. \ref{intro} we introduced the empirical IRX-$\beta$ relation, that is used to correct for dust obscuration in UV-selected galaxies at $z>5$. This relation is confirmed to hold up to $z\sim 2-3$, but it is yet unclear whether it applies at higher redshift where a large scatter is present in the data \citep{barisic2017dust,fudamoto2017dust}. On average, high-$z$ galaxies are shown to have low IRX, and lie below the local relation \citep{capak15,Bouwens16}. However, as discussed in Sec. \ref{intro}, this conclusion is heavily dependent on the dust temperature assumption for the SEDs of dust emission, which is generally set in $T_{d}=35-45\ \mathrm{K}$. This might lead to a severe underestimate of the high-$z$ galaxies FIR luminosity (see the re-analysis of the sample by \citealt{capak15} performed by \citealt{barisic2017dust})
 
As already pointed out by \cite{Bouwens16}, the tension with the local relation would be alleviated by warmer dust, with $T_{\rm d}=45-50\ \mathrm{K}$. In addition, even if the bulk of the dust remained cold ($\sim 35\ \mathrm{K}$), a moderate fraction of warm dust is already sufficient to affect the SEDs (see the discussion in Sec. \ref{masswelumwe}), reducing the apparent flux on the Rayleigh-Jeans tail at fixed $L_{\rm FIR}$ \citep{casey2018}. This scenario is consistent with our conclusion on dust temperature in high-$z$ GMCs, where due to the large turbulent velocities and consequent high SFR, dust attains temperatures as high as $T_{\rm d} \simeq 60\ \mathrm{K}$. The resulting increase in FIR luminosity, and consequently in IRX parameter, is as large as $1.4\ \mathrm{dex}$ (computed using eq. \ref{Tdp}) with respect to considering $T_{\rm d}=35\ \mathrm{K}$. This is sufficient to shift the IRX faint galaxies back on the local relation. 
As a final remark, we notice that IRX-$\beta$ relation implicitly assumes  that FIR and UV emitting regions are co-spatial. However, if dust completely absorbs the radiation emitted by stars within GMCs, as suggested here, FIR and UV emitting regions might actually be spatially segregated \citep{behrens2018dusty}.

\subsection{Dust mass estimates from $L_{\rm FIR}$}\label{dustmass}

Higher dust temperatures imply that a smaller dust amount can produce the same FIR luminosity, see eq. \ref{Tdp}. This would alleviate the problem of a super-efficient dust production by stellar sources which is very hard to reconcile with available data and theoretical results; see discussion in e.g. \cite{2019ApJ...874...27T} for the galaxy MACS0416Y1 at $z=8.31$ (whose $90\ \mathrm{\mu m}$ continuum is observed). 

For instance given $L_{\rm FIR}$, assuming $T_{\rm d}=60\ \mathrm{K}$, which is the maximum value derived for our high-$z$ GMC (when the radiation pressure effect is included), would result in a dust mass reduction by a factor $\Delta M_{\rm d} = (50/60)^{4+\beta_{\rm d}}=0.33$ with respect to the value suggested by \cite{2019ApJ...874...27T} who took $T_{\rm d}=50\ \mathrm{K}$. The power $4+\beta_{\rm d}=6$ is derived from the integration of the greybody function (eq. \ref{Tdp}). 
\begin{figure}
    \centering
    \includegraphics[width=1.0\linewidth]{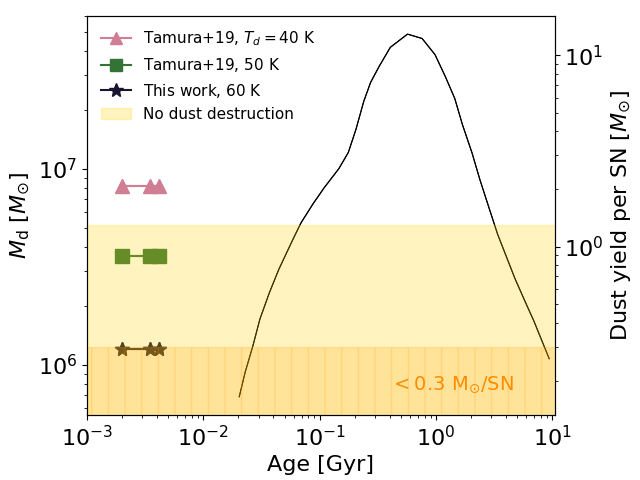}
    \caption{Time evolution of dust mass predicted in a dust formation model \citep[e.g.][]{asano13,nozawa2014evolution} as a function of galaxy age with an initial gas mass of $M_{\rm gas}= 2 \times 10^{10}\ \msun$ and a star formation time scale of $\tau_{\rm SFH}= 0.3\ \mathrm{Gyr}$. This plot is adapted from \citet[][ Fig. 6, lower right panel]{2019ApJ...874...27T}. The symbols correspond to the best-fitting parameters of the physical properties of MACS0416Y1 ($z=8.31$) estimated through \citet{calzetti2000dust}, MW, and SMC extinction laws (from right to left). Each colour and shape corresponds to a different assumption for the dust temperature $T_{\rm d}$, as shown by the legend. On the left y-axis we also show which would be the dust yield per SN event depending on the computed dust mass, and so on the assumed dust temperature (see text). The yellow shaded region corresponds to the maximum theoretical dust yield without dust destruction in SNe explosions, and the hatched orange region to the upper limit assuming $T_{\rm d}=60\ \mathrm{K}$.
    \label{figtamura}
    }
\end{figure}
This result is shown graphically in Fig. \ref{figtamura}, where we compare the dust mass estimates for MACS0416Y1 resulting from three different assumptions for the dust temperature, $T_{d}=(40,50,60)\ \mathrm{K}$. 

We can also compute the dust mass per SN event produced for these three cases, by using the stellar mass computed by \cite{2019ApJ...874...27T}, $M_{*}\sim 2 \times 10^{8}\ \mathrm{M_{\odot}}$:
\begin{equation}
    \mathrm{Dust\ yield\ per\ SN} = \frac{M_{\rm d}}{M_{*}\nu_{\rm SN}}\, , 
\end{equation}
where $\nu_{\rm SN}$ (see eq. \ref{tSN}) is the number of SNe per solar mass of stars formed. We obtain that for $T_{d}= 50\ \mathrm{K}$, which is the most optimistic case considered by \cite{2019ApJ...874...27T}, around $\sim 1\ \mathrm{M_\odot}$ of dust should be produced in each SNe explosion. This value exceeds the most recent constraints on SN dust yields by \cite{lesniewska2019dust} when we allow for some dust destruction during SNe explosions. Moreover, for the most frequently assumed value of the dust temperature $T_{\rm d}= 40\ \mathrm{K}$, we get $2\ \mathrm{M_{\odot}}$ of dust yield per SN event, which exceeds even the less stringent constraints available (i.e. with no dust destruction, see the yellow shaded region in Fig. \ref{figtamura}). Instead, if we consider $T_{\rm d}= 60\ \mathrm{K}$, dust yield drops to a modest $0.3\ \mathrm{M_{\odot}}$, a value more in line with current data.

We also note that recent observations of MACS0416Y1, indicate the presence of warmer dust in this galaxy. \cite{bakx20} fail to detect continuum emission at $160\ \mathrm{\mu m}$ (rest frame) down to $18\ \mathrm{\mu Jy}$. This non-detection places strong limits on the dust spectrum, given the observed continuum emission at longer wavelengths $850\ \mathrm{\mu m}$ \citep{2019ApJ...874...27T}. In particular, it suggests an unusually warm dust component $T_{\rm d,L}>80\ \mathrm{K}$ ($90\%$ confidence  limit) considering an emissivity index $\beta_{\rm d}=2$. This would further decreases the required dust masses, with a reduction factor of $0.06$ with respect to the value deduced by \cite{2019ApJ...874...27T}.

\section{Summary}\label{summary}

In this work, we have developed a detailed model for dust temperature and associated Far Infrared (FIR) emission within star forming Giant Molecular Clouds (GMC). We emphasise the difference between local and high-$z$ GMCs, which is connected mainly to the larger pressure ($p$) and the turbulent velocity ($\sigma$) within early galaxies. These differences have two crucial effects: the star formation rate, which we showed to depend only on $\sigma^{3}$, is $\sim 30$ times larger in high-$z$ clouds, and the optical depth to Ultra Violet (UV) is $\sim 10$ times higher due to the increase in the cloud column density, which depends on ${p}^{1/2}$. We have also analysed the effects of UV radiation for the formation of \HII regions and radiation pressure -- in short radiative feedback -- which, along with mechanical feedback from stellar winds and SNe, determine the cloud dispersal time scale.

Our main findings are summarised as follows: 
\begin{itemize}
    \item \textbf{Dust temperature in star forming GMCs}: in high-$z$ clouds dust is warmer than in local ones, reaching $T_{\rm d}=60\ \mathrm{K}$. This results from (i) the more compact structure of GMCs, and (ii) the more turbulent nature of early galaxies;
    \item \textbf{Clouds lifetimes}: star formation (and associated UV emission) is heavily obscured within the very compact high-$z$ clouds. Due to their compact nature, these structures also survive longer (about $\sim 4 \times$) than local clouds to \HII region expansion, stellar winds and SNe. Their lifetimes are $\sim 10$ Myr. 
    \item \textbf{Infrared emission from star forming GMCs}: high-$z$ clouds are intense FIR emitters, with a luminosity-to-dust mass ratio of $\sim 10\  \mathrm{L_{\odot}/M_{\odot}}$, which is 10 times higher than the value of our reference MW cloud.
\end{itemize}

These results have some very significant implications. The unobscured UV emission observed in high-$z$ galaxies mainly comes from clouds that have already been dispersed, and hence depends on cloud lifetimes. This may lead to a spatial separation between the UV, which arises from the diffuse ISM, and the FIR emission, associated to compact clouds. Our scenario seems to be supported by simulations \citep{behrens2018dusty} and observations \citep[e.g. ][]{laporte:2017apj,bowler2018obscured}, which find a significant spatial offset between ALMA and HST data. The spatial offset between UV and FIR might lead to questioning the significance of the IRX-$\beta$ relation at high redshift.
In any case, the higher dust temperatures predicted here, would reduce the tension between local and high-$z$ IRX-$\beta$ relation, shifting the early IRX faint galaxies back on the local relation. 
Warmer dust also reduces the problem of insufficient dust production due to star formation in early times ($<0.1\ \mathrm{Gyr}$). Indeed, the dust yield produced per SN event within our clouds is in agreement with the most updated SN efficiency constraints. 

Finally, warmer dust produces a strong interstellar FIR radiation field pervading the galaxy, and hence GMCs. This \quotes{background} might affect FIR emission lines, such as the fine-structure \CII $158\ \mathrm{\mu m}$ line in two ways by (1) altering the population levels of $C^{+}$, and (2) decreasing the line-to-continuum contrast of \CII. These two effects might possibly be responsible for the observed \CII deficit in luminous FIR emitting galaxies ($L_{\rm FIR}>10^{11}\ \mathrm{L_{\odot}}$). In addition it could induce  significant spatial variations in the \CII/FIR ratio as hinted by recent observations of high-$z$ starbusts \citep[e.g.][]{2016ApJ...827...34O}. 

Given the relevant implications that warm/hot dust may have for galaxy formation at early epochs, it is of utmost importance to measure dust temperature in high-$z$ galaxies. This will become possible in the near future by means of planned infrared observatories such as SPICA \citep[e.g.][]{spinoglio:2017,egami:2018}.

\section*{Acknowledgements}
LS, AF, SC, and DD acknowledge support from the ERC Advanced Grant INTERSTELLAR H2020/740120. Any dissemination of results must indicate that it reflects only the author's view and that the Commission is not responsible for any use that may be made of the information it contains.
Partial support (AF) from the Carl Friedrich von Siemens-Forschungspreis der Alexander von Humboldt-Stiftung Research Award is kindly acknowledged.
We thank Rebecca Bowler, Alessandro Lupi, and Pascal Oesch for useful discussions.
We acknowledge use of the Python programming language \citep{VanRossum1991}, Matplotlib \citep{Hunter2007}, Numpy \citep{VanDerWalt2011} and Scipy \citep{scipy2019}.
% Astropy \citep{astropy}, Cython \citep{behnel2010cython},

\bibliographystyle{mnras}
\bibliography{accepted_paper}

% Don't change these lines
\bsp	% typesetting comment

\label{lastpage}
\end{document}